# The Cambridge-Cambridge *ROSAT* Serendipity Survey – I. X-ray-luminous galaxies

B.J. Boyle,[1,2] R.G. McMahon,[1] B.J. Wilkes[3] and M. Elvis[3]

[1] *Institute of Astronomy, University of Cambridge, Madingley Road, Cambridge, CB3 0HA*
[2] *Present address: Royal Greenwich Observatory, Madingley Road, Cambridge, CB3 0EZ*
[3] *Center for Astrophysics, 60 Garden St, Cambridge, MA 21260, USA*

**ABSTRACT**
We report on the first results obtained from a new optical identification programme of 123 faint X-ray sources with $S(0.5–2\,\mathrm{keV}) > 2 \times 10^{-14}\,\mathrm{erg\,s^{-1}\,cm^{-2}}$ serendipitously detected in *ROSAT* PSPC pointed observations. We have spectroscopically identified the optical counterparts to more than 100 sources in this survey. Although the majority of the sample (68 objects) are QSOs, we have also identified 12 narrow emission line galaxies which have extreme X-ray luminosities ($10^{42} < L_\mathrm{X} < 10^{43.5}\,\mathrm{erg\,s^{-1}}$). Subsequent spectroscopy reveals them to be a mixture of starburst galaxies and Seyfert 2 galaxies in approximately equal numbers. Combined with potentially similar objects identified in the *Einstein* Extended Medium Sensitivity Survey, these X-ray-luminous galaxies exhibit a rate of cosmological evolution, $L_\mathrm{X} \propto (1 + z)^{2.5 \pm 1.0}$, consistent with that derived for X-ray QSOs. This evolution, coupled with the steep slope determined for the faint end of the X-ray luminosity function ($\Phi(L_\mathrm{X}) \propto L_\mathrm{X}^{-1.9}$), implies that such objects could comprise 15–35 per cent of the soft (1–2 keV) X-ray background.

**Key words:** X-rays: general – galaxies: active – quasars: general

## 1 INTRODUCTION

The launch of *ROSAT* has provided a new opportunity to resolve the soft (0.5–2 keV) X-ray background (XRB) down to flux levels five times fainter than achieved with the deepest pointings performed by the *Einstein* satellite (Griffiths et al. 1992). At these levels ($S \sim 2 \times 10^{-15}\,\mathrm{erg\,s^{-1}\,cm^{-2}}$) almost 60 per cent of the XRB is resolved (Hasinger et al. 1993) and, at the current limit of the follow-up optical identification programmes ($S \sim 4 \times 10^{-15}\,\mathrm{erg\,s^{-1}\,cm^{-2}}$, Georgantopoulos et al. 1995), QSOs account for the majority ($\sim 75$ per cent) of the extragalactic sources. However, fluctuation analyses (Hasinger et al. 1993; Georgantopoulos et al. 1993) and differences between the X-ray spectra of QSOs and that of the background (Georgantopoulos et al. 1995) suggest that these QSOs are unlikely to contribute more than $\sim 50$ per cent of the 1-keV background and probably do not comprise more than $\sim 20$ per cent of the XRB at harder energies (2–10 keV), the energy range at which the XRB was first discovered (Giacconi et al. 1962). Indeed, at faint flux levels, $S < 10^{-14}\,\mathrm{erg\,s^{-1}\,cm^{-2}}$, the spectroscopic observations of Georgantopoulos et al. (1995) tentatively identify an increasingly large fraction of galaxies as counterparts to X-ray sources.

Unfortunately the low flux levels currently employed in both the X-ray and optical regimes make optical identification difficult and positional coincidence hard to establish conclusively (e.g. Georgantopoulos et al. 1995). By slightly increasing the flux limit and extending any survey over a wider area, it may be possible to identify the brighter members of any new class. We have therefore carried out an optical identification programme of X-ray sources detected by *ROSAT* at a level intermediate in flux, $S(0.5 - 2\,\mathrm{keV}) = 2 \times 10^{-14}\,\mathrm{erg\,s^{-1}\,cm^{-2}}$, between the deepest *ROSAT* surveys and the Extended Medium Sensitivity Survey (EMSS, see Stocke et al. 1991) performed with the *Einstein* satellite at a flux limit $S(0.3 - 3.5\,\mathrm{keV}) \sim 1 \times 10^{-13}\,\mathrm{erg\,s^{-1}\,cm^{-2}}$. The sources in our survey were detected 'serendipitously' in existing PSPC fields available in the *ROSAT* archive, originally taken to study the X-ray properties of known targets (QSOs/CVs/galaxies). The optical counterparts of these X-ray sources were identified spectroscopically with the William Herschel Telescope (WHT). A full description of the survey and catalogue will appear elsewhere (McMahon et al., in preparation, hereafter M95). In this paper we report on the discovery of a significant number of narrow emission line galaxies in the survey. In Section 2 we briefly describe the selection of the X-ray sources, the optical identification procedure and the spectroscopic observations. We discuss our results in Section 3 and present our conclusions in Section 4.

**Table 1.** Survey fields.

| Field Name | RA (J2000) Dec | | | Flux Limit | $N_H$ |
|---|---|---|---|---|---|
| | h  m   s | °   ′   ″ | | (erg s$^{-1}$ cm$^{-2}$) | (10$^{20}$ cm$^{-2}$) |
| MKN335 | 00 08 19.2 | 20 41 24 | | $2.0 \times 10^{-14}$ | 4.1 |
| PG0027 | 00 30 05.7 | 26 17 22 | | $2.0 \times 10^{-14}$ | 3.9 |
| OQ208 | 14 07 00.4 | 28 27 14 | | $2.0 \times 10^{-14}$ | 1.4 |
| PG1411 | 14 13 48.4 | 44 00 14 | | $2.0 \times 10^{-14}$ | 1.2 |
| Q1413+11 | 14 15 46.3 | 11 29 44 | | $2.0 \times 10^{-14}$ | 1.8 |
| 3C298 | 14 19 08.2 | 06 28 35 | | $2.0 \times 10^{-13}$ | 2.1 |
| OQ530 | 14 19 46.6 | 54 23 14 | | $2.0 \times 10^{-14}$ | 1.3 |
| PG1426 | 14 29 06.6 | 01 17 06 | | $2.5 \times 10^{-14}$ | 2.7 |
| 4U1417+42 | 14 28 32.5 | 42 40 25 | | $3.0 \times 10^{-14}$ | 1.4 |
| PG1512 | 15 14 43.0 | 36 50 51 | | $4.0 \times 10^{-14}$ | 1.4 |
| NGC5907 | 15 15 52.9 | 56 19 33 | | $2.0 \times 10^{-14}$ | 1.4 |
| PG1543 | 15 45 30.1 | 48 46 12 | | $3.0 \times 10^{-14}$ | 1.6 |
| MS1603.6 | 16 05 46.0 | 25 51 44 | | $2.0 \times 10^{-14}$ | 4.6 |
| E1615+061 | 16 17 45.5 | 06 03 55 | | $3.5 \times 10^{-14}$ | 4.6 |
| 3C334 | 16 20 21.8 | 17 36 22 | | $5.0 \times 10^{-14}$ | 4.1 |
| MARK501 | 16 53 52.2 | 39 45 36 | | $7.5 \times 10^{-14}$ | 1.7 |
| PG1704 | 17 04 41.5 | 60 44 28 | | $2.0 \times 10^{-14}$ | 2.3 |
| PG2233 | 22 36 07.7 | 13 43 55 | | $5.0 \times 10^{-14}$ | 4.9 |
| PKS2247 | 22 50 25.3 | 14 19 53 | | $4.0 \times 10^{-14}$ | 4.8 |
| HR8905 | 23 25 22.1 | 23 24 13 | | $3.0 \times 10^{-14}$ | 4.7 |

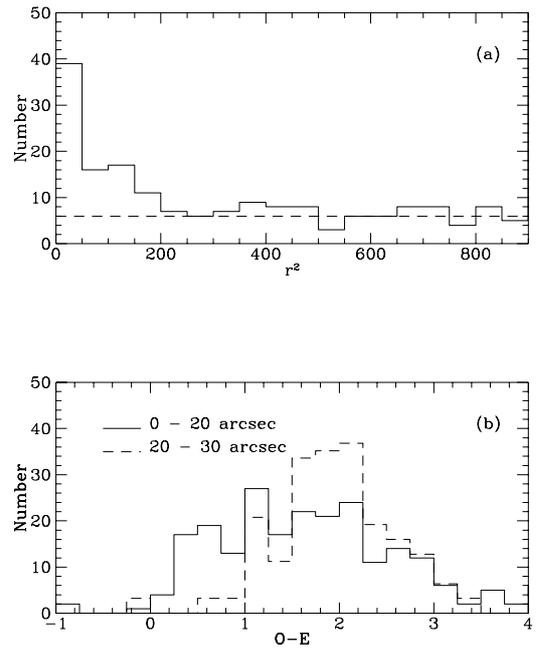

**Figure 1.** (a) Number of optical identifications as a function of separation from X-ray source position. Bins are in units of separation$^2$ (arcsec$^2$) so the level of chance coincidences is constant (dashed line). (b) $O - E$ colour distributions for the optical counterparts within 20 arcsec of the X-ray position, and the overall colour distribution of the field, based on the normalized distribution of colours for optical identifications lying $20 - 30$ arcsec from the X-ray sources.

## 2  OBSERVATIONS

### 2.1  X-ray source detection

Objects in the Cambridge-Cambridge *ROSAT* serendipity survey (CRSS) were detected as serendipitous X-ray sources in high galactic latitude ($b > 30°$) PSPC fields available in the *ROSAT* archive prior to 1993 July. All selected fields had exposure times greater than 6000 s. X-ray sources were detected on 20 PSPC fields with $14^h < RA < 1^h$ and $\delta > 0°$. On each field source detection was limited to within 15 arcmin off-axis in order to maintain relatively high positional accuracy for the sources ($1\sigma$ error $\simeq 8$ arcsec) and to avoid obscuration by the rib support structure at larger off-axis angles. Sources were detected using the IMAGES algorithm written by Dr Mike Irwin (now available as PISA in the STARLINK software collection), which detected sources with four or more 15-arcsec pixels lying 1.5 times the rms noise level above the background. Source detection on each field was limited to a flux level $S(0.5-2 \text{ keV}) > 2 \times 10^{-14} \text{ erg s}^{-1} \text{ cm}^{-2}$, significantly greater than the faintest flux detectable even on the fields with the shortest exposures. Source fluxes were measured using a fixed 2.5-pixel radius aperture. Background subtraction was made using a single level for the whole field (typically $< 0.3$ count pixel$^{-1}$), determined from a fit to the histogram of pixel values within the 15-arcmin radius region. On more than half of the fields in this analysis (11), the spectroscopic identification is 90 per cent complete to the flux limit $S(0.5 - 2 \text{ keV}) > 2 \times 10^{-14} \text{ erg s}^{-1} \text{ cm}^{-2}$. On the other fields (principally the shorter exposure fields), we subsequently imposed a brighter flux limit to ensure a similar level of spectroscopic completeness on these fields. The revised area coverage as a function of X-ray flux limit in the $0.5 - 2$ keV passband can be derived from Table 1. A total of

123 serendipitous sources were identified over the 20 PSPC fields used in this analysis.

### 2.2  Optical identification

In order to search for optical counterparts, the measured $X, Y$ positions for the X-ray sources were transformed to RA and Dec. using their offsets from the central 'target' source in each PSPC field (usually a QSO) whose optical position was known to 1-arcsec accuracy. APM measurements of Palomar Observatory Sky Survey (POSS) O and E plates (McMahon & Irwin, in preparation) were then used to provide a list of optical counterparts found within 20 arcsec of each source. Out to 20 arcsec, 112 out of 123 X-ray sources (91 per cent) had at least one optical counterpart identified to the limit of the POSS plates ($O < 21.5$ mag, $E < 20$ mag). The 20-arcsec radius represents 2.5 times the observed width of the X-ray to optical positional cross-correlation function (Fig. 1(a)). The mean $O - E$ colour of these counterparts is much bluer than the overall colour distribution (Fig. 1(b)), demonstrating that the counterparts are dominated by QSOs ($O - E \sim 1$ mag, McMahon 1991). The X-ray source positions were subsequently revised to give a zero mean positional offset in each field between the X-ray and optical positions of sources with a blue ($O-E < 1.2$ mag) optical counterpart. However, this changed the nearest optical counterpart in only 5 per cent of the total sample. As can be seen from Fig. 1(a), a significant number of optical counterparts ($\sim 35$ per cent) identified within 20 arcsec will be chance coincidences, giving rise to multiple optical counterparts for some X-ray sources. For sources with more than one optical counterpart, spectroscopy was carried out in strict order of increasing separation between the X-ray position and optical counterpart until a plausible candidate (e.g. QSO, M star) was identified. In all but 10 cases the closest object turned out to be the correct identification.

## 2.3 Optical spectroscopy

Low-resolution ($\sim 6$ Å) spectroscopy was obtained for the optical counterparts with the ISIS double arm spectrograph at the WHT on the nights of 1993 June 19–22. ISIS was operated with 158 line mm$^{-1}$ gratings in the blue and red arms (a dichroic filter was used to split the light at 5400 Å), with the Tektronix and EEV CCDs used as the detectors in the blue and red arms respectively, giving an overall instrumental resolution of 2.9 Å pixel$^{-1}$. Throughout the run the seeing was $\sim 1$ arcsec, and most of the observations were made with a 1.5-arcsec slit. In these conditions, exposure times of 300–1200 s were sufficient to yield an adequate signal-to-noise ratio ($\sim 15$) in the spectra of the optical candidates ($19 < O < 21$). Spectra for 118 optical candidates were obtained, resulting in a positive optical identification for 107 of the 123 X-ray sources in the CRSS (including three previously known QSOs). An additional two bright ($O < 12$ mag) stellar optical counterparts were not observed, but, if they are assumed to be galactic stars, the identification rate for the survey is 89 per cent. Of the remaining 14 unidentified X-ray sources, 11 had no optical counterpart within 20 arcsec of the X-ray position on the POSS, one optical counterpart had an inconclusive, low signal-to-noise ratio spectrum, and a further optical counterpart was not observed due to lack of time. A Kolmogorov-Smirnov (KS) test performed on the X-ray flux distributions of both the identified and unidentified sources revealed no significant differences between the flux distributions for the two samples at the 95 per cent confidence level. To account for this incompleteness, in the analysis below we have therefore multiplied the total area surveyed at each X-ray flux limit by 0.89 to yield an effective survey area.

# 3 X-RAY GALAXIES

## 3.1 General properties

The majority of the objects identified were QSOs (68), classified from the presence of broad ($> 1000$ km s$^{-1}$) emission lines. Full details of the properties of the QSO population (e.g. evolution, luminosity function) will be discussed by M95.

We also identified 12 narrow emission line X-ray galaxies (NLXGs) with optical magnitudes in the range $17 < O < 21$. The criterion for inclusion in this class was a measured H$\alpha$ full width half maximum (FWHM) of $< 1000$ km s$^{-1}$. The limit clearly marks the division between broad and narrow emission lines in the CRSS survey (see Fig. 2). The discovery spectra for all these galaxies are plotted in Fig. 3 and their names, positions, redshifts, $0.5 - 2$ keV fluxes and luminosities are given in Table 2. Luminosities were derived assuming $q_0 = 0.5$, $H_0 = 50$ km s$^{-1}$ Mpc$^{-1}$ and an X-ray spectral index $\alpha_X = 1.0$. The X-ray luminosities for these galaxies lie in the range $10^{42} < L_X < 10^{43.5}$ erg s$^{-1}$, one to two orders of magnitude more luminous than late-type galaxies (Fabbiano 1989), which these galaxies most closely resemble in terms of their optical spectra. All the X-ray images for these galaxies are consistent with a point source, ruling out any significant contribution to the X-ray luminosity from extended cluster emission at the typical redshift ($z \sim 0.2$) of the sample.

It is unlikely that galaxies have been identified merely by chance positional coincidence. All galaxies were found within 10 arcsec of the X-ray source position. Based on the mean surface density of optical counterparts to the plate limit (see Fig. 1(a)) we expect only 15 per cent of the X-ray sources to have a spurious optical counterpart at this separation. Of these chance coincidences, approximately half will be galaxies (based on the relative numbers of stars and galaxies identified at the POSS plate limit) and, of these galaxies, $\sim 20 - 30$ per cent will exhibit emission lines (based on the relative number of emission line galaxies at this magnitude limit, Broadhurst, Ellis & Shanks 1988). Thus only $\sim 1 - 2$ per cent of the total sample ($\sim 1 - 2$ X-ray sources) will have a positional coincidence (at $< 10$ arcsec) with an emission line galaxy. An even smaller fraction will exhibit the high [OIII]/[OII] emission line ratios observed for galaxies in this sample (see below), and thus we consider it likely that the vast majority, if not all, of these galaxies are the true counterparts of the X-ray sources.

The galaxies exhibit emission lines of [OII]$\lambda$3727, H$\beta$, [OIII]$\lambda\lambda$4959, 5007, [NII]$\lambda\lambda$6548, 6583, H$\alpha$ and [S II]$\lambda\lambda$6716, 6731, typical of many late-type galaxies identified in field galaxy surveys at this optical magnitude limit (Broadhurst et al. 1988). However, the observed emission line ratios, in particular the strength of [OIII] relative to [OII] ($1 <$ [OIII]$\lambda$5007/[OII]$\lambda$3727 $< 4$), indicate a much higher state of ionization than is normally observed for field galaxies.

We have managed to obtain a more accurate classification for 10 of the NLXGs using subsequent higher resolution (1.5 Å), higher signal-to-noise ratio ISIS spectra of the redshifted H$\alpha$ and H$\beta$ regions in these objects. These spectra were obtained during an observing run on 1994 June 9-10 and during service time on 1993 September 20. A detailed discussion of these spectra will be presented elsewhere (Boyle et al., in preparation), but a classification based on the relative strengths of the H$\beta$, [OIII]$\lambda$5007, [OI]$\lambda$6300, H$\alpha$ and [NII]$\lambda$6583 lines using the schemes of Baldwin, Philips & Terlevich (1981) and Filippenko & Terlevich (1992) reveals that the sample is composed of approximately equal numbers of starburst-like galaxies and Seyfert 2-like galaxies, although many objects lie close to the transition regions in the classification schemes used. Emission line ratios and classifications for these objects are presented in Table 2. The measured FWHM for the Balmer lines in these spectra lie in the range $180$ km s$^{-1}$ < FWHM $< 450$ km s$^{-1}$, confirming their original identification as narrow-lined objects, and distinct from broad-lined QSOs. However, as many as half the sample may exhibit weak broad emission (FWHM $> 1000$ km s$^{-1}$) at the base of H$\alpha$, although this is very much dependent on the adopted emission line profile used to fit the more prominent narrow emission lines. In only one case (CRSS1413.3+4405) is there clear evidence for a residual weak broad component (FWHM$\sim$ 6000 km s$^{-1}$), even after logarithmic line profiles have been used to fit the narrow emission lines.

**Table 2.** NLXGs identified in the CRSS.

| Name | Survey Name | RA (2000) Dec | | | | | | $S(0.5-2\,\mathrm{keV})$ | z | $L(0.5-2\,\mathrm{keV})$ | $\frac{\mathrm{I([OIII])}}{\mathrm{I(H\beta)}}$ | $\frac{\mathrm{I([NII])}}{\mathrm{I(H\alpha)}}$ | ID |
|------|-------------|---|---|---|---|---|---|---|---|---|---|---|---|
| | | h | m | s | ° | ′ | ″ | $(\mathrm{erg\,s^{-1}\,cm^{-2}})$ | | $(\mathrm{erg\,s^{-1}})$ | | | |
| CRSS0009.0+2041 | MKN335:08 | 00 | 09 | 01.7 | 20 | 41 | 36 | $2.0\times10^{-14}$ | 0.189 | $3.4\times10^{42}$ | | 0.5 | HII? |
| CRSS0030.2+2611 | PG0027:10 | 00 | 30 | 17.4 | 26 | 11 | 38 | $5.1\times10^{-14}$ | 0.077 | $1.7\times10^{42}$ | 0.2 | 0.5 | HII |
| CRSS0030.7+2629 | PG0027:19 | 00 | 30 | 47.9 | 26 | 29 | 34 | $2.0\times10^{-14}$ | 0.246 | $5.9\times10^{42}$ | | 0.9 | Sy 2 |
| CRSS1406.7+2838 | OQ208:43 | 14 | 06 | 47.9 | 28 | 38 | 53 | $2.8\times10^{-14}$ | 0.331 | $1.5\times10^{43}$ | | 0.3 | HII? |
| CRSS1412.5+4355 | PG1411:14 | 14 | 12 | 31.6 | 43 | 55 | 36 | $2.2\times10^{-13}$ | 0.094 | $8.9\times10^{42}$ | 3.6 | 0.7 | Sy 2 |
| CRSS1413.3+4405 | PG1411:22 | 14 | 13 | 19.9 | 44 | 05 | 34 | $5.2\times10^{-14}$ | 0.136 | $4.5\times10^{42}$ | 7.6 | 0.8 | Sy 2 |
| CRSS1415.0+4402 | PG1411:19 | 14 | 15 | 00.1 | 44 | 02 | 08 | $2.9\times10^{-14}$ | 0.136 | $2.5\times10^{42}$ | 7.9 | 0.8 | Sy 2 |
| CRSS1429.0+0120 | PG1426:03 | 14 | 29 | 04.7 | 01 | 20 | 17 | $4.3\times10^{-14}$ | 0.102 | $2.0\times10^{42}$ | 1.5 | 0.6 | Sy 2 |
| CRSS1514.4+5627 | N5907:15 | 15 | 14 | 29.6 | 56 | 27 | 07 | $3.7\times10^{-14}$ | 0.446 | $3.8\times10^{43}$ | | | – |
| CRSS1605.6+2543 | MS1603:07 | 16 | 05 | 39.9 | 25 | 43 | 10 | $4.4\times10^{-14}$ | 0.278 | $1.7\times10^{43}$ | 0.5 | 0.4 | HII |
| CRSS1605.9+2554 | MS1603:16 | 16 | 05 | 58.7 | 25 | 54 | 08 | $4.1\times10^{-14}$ | 0.151 | $4.4\times10^{42}$ | | | – |
| CRSS1705.3+6049 | PG1704:11 | 17 | 05 | 18.3 | 60 | 49 | 54 | $2.7\times10^{-14}$ | 0.572 | $4.6\times10^{43}$ | > 2 | | Sy 2? |

**Table 3.** NLXGs observed in the EMSS.

| Name | z | $L(0.3-3.5\,\mathrm{keV})$ $(\mathrm{erg\,s^{-1}})$ | $\frac{\mathrm{I([OIII])}}{\mathrm{I(H\beta)}}$ | $\frac{\mathrm{I([NII])}}{\mathrm{I(H\alpha)}}$ | ID |
|------|---|---|---|---|---|
| MS1252.4+0457 | 0.158 | $2.6\times10^{43}$ | 0.58 | 0.88 | Sy 2 |
| MS1334.6+0351 | 0.136 | $2.4\times10^{43}$ | 3.30 | 1.05 | Sy 2 |
| MS1412.8+1320 | 0.139 | $2.2\times10^{43}$ | 0.58 | 0.36 | HII |
| MS1414.8−1247 | 0.198 | $1.3\times10^{44}$ | 2.08 | 0.33 | HII |
| MS1555.1+4522 | 0.181 | $6.9\times10^{43}$ | 7.05 | 0.91 | Sy 2 |
| MS1614.1+3239 | 0.118 | $1.7\times10^{43}$ | > 10 | 1.23 | Sy 2 |
| MS2044.1+3532 | 0.183 | $3.7\times10^{43}$ | 2.7 | 0.52 | Sy 2/HII |

Similar examples of X-ray-luminous ($10^{42} < L_X < 10^{43}\,\mathrm{erg\,s^{-1}}$) narrow emission line galaxies were also identified at low redshift ($z \lesssim 0.2$) in the EMSS by Stocke et al. (1991). Since the emission line galaxies fulfilled the criterion that [OIII]$\lambda$5007 >[OII]$\lambda$3727, they were classified as AGN without exhibiting any broad emission lines. Based on the list of 'ambiguous' sources listed in tables 8 and 10 of Stocke et al. (1991), it is possible that up to 31 objects classified as AGN in the EMSS could be similar to the types of NLXGs identified in the CRSS. We observed 7 of these objects with ISIS at 1.5-Å resolution during the WHT run on 1994 June 9-10 and found that the FWHM of the Hα and Hβ lines lay in the range 200–600 km s$^{-1}$, although, as was the case for the CRSS objects, the presence of weak broad lines was difficult to establish due to uncertainties in fitting profiles to the wings of the narrow emission lines. The measured emission line ratios and corresponding classifications for these seven EMSS objects are given in Table 3. Despite the small numbers of objects involved, the relative numbers of starburst-like spectra and Seyfert 2-like spectra seem approximately similar to those identified in the CRSS. Moreover, these numbers agree with those found from a similar spectroscopic classification of the ambiguous EMSS sources performed by Fruscione, Griffiths & Mackenty (1993). Although it would appear unlikely that the NLXGs in either the CRSS or EMSS form a homogeneous group (although most lie close to the classification boundary between starburst galaxies and Seyfert 2s), an overall estimate of their total space density and evolution can be used to provide a useful upper limit to the contribution of NLXGs (starburst and Seyfert 2) to the 1–2 keV XRB. We have therefore combined the EMSS 'ambiguous' sources with the CRSS NLXGs to give a total sample of 43 objects, which will be used as the basis for the analysis below.

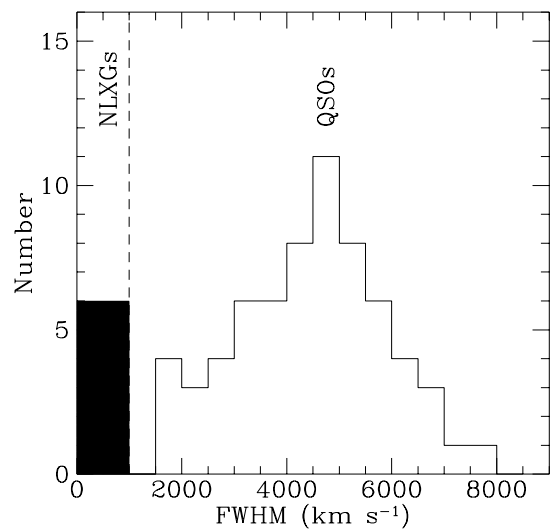

**Figure 2.** Histogram of emission line full width half maxima (FWHM) for the QSOs and NLXGs in the CRSS. The classification criterion used to identify NLXGs (FWHM < 1000 km s$^{-1}$) is indicated by the dashed line, and the NLXGs are denoted by the shaded region.

### 3.2 Evolution

The NLXG fraction in the total X-ray source counts (including non-extragalactic sources) increases with flux limit (4 per cent in the EMSS, $S(0.3-3.5\,\mathrm{keV}) > 1 \times$ $10^{-13}\,\mathrm{erg\,s^{-1}\,cm^{-2}}$; and 10 per cent in the CRSS, $S(0.5-2\,\mathrm{keV}) > 2 \times 10^{-14}\,\mathrm{erg\,s^{-1}\,cm^{-2}}$) in a similar manner to QSOs. To establish whether this is due to cosmological evolution, we calculated the $< V_e/V_a >$ statistic (Avni & Bahcall 1980) for the 43 NLXGs in the EMSS and CRSS. We first converted the 0.5–2 keV *ROSAT* fluxes to the *Einstein* band assuming $S(0.3-3.5\,\mathrm{keV}) = 1.8 \times S(0.5-2\,\mathrm{keV})$, appropriate for an X-ray spectral index $0.8 < \alpha_X < 1.5$ (Boyle et al. 1993). For $q_0 = 0.5$ we obtained a value of $< V_e/V_a > = 0.72 \pm 0.04$, more than $5\sigma$ higher than the no-evolution value of 0.5.

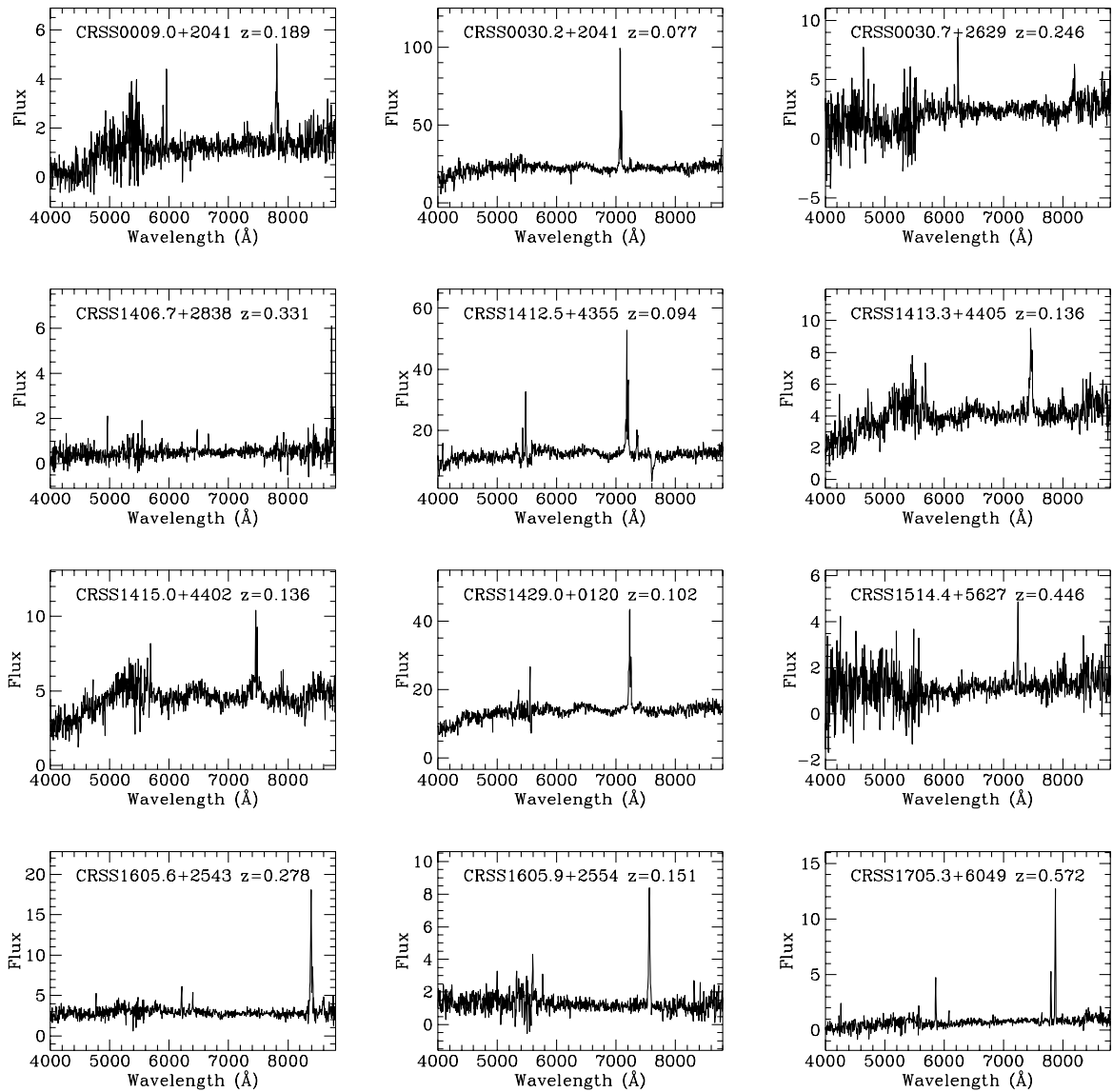

**Figure 3.** Low-resolution (10 Å) spectra for all 12 narrow emission line galaxies identified in the CRSS. The flux scale is plotted in arbitrary units.

Having confirmed that the NLXG sample exhibits significant cosmological evolution, we then used the maximum likelihood technique to obtain a 'best-fit' parametric representation of the evolution and luminosity function (LF) of this sample. We adopted a two-power-law model for the NLXG LF, also found to fit the QSO X-ray LF (Boyle et al. 1993):

$$\Phi(L_X) = \Phi^* L_{X_{44}}^{-\gamma_1} \qquad L_X < L_X^*(z=0)$$

$$\Phi(L_X) = \frac{\Phi^*}{L_{X_{44}}^{\gamma_1-\gamma_2}} L_{X_{44}}^{-\gamma_2} \qquad L_X > L_X^*(z=0)$$

where $\Phi^*$ is the normalization of the LF and $\gamma_1$, $\gamma_2$ are the slopes of the faint and bright ends of the LF. $L_{X_{44}}$ is the 0.3–3.5 keV X-ray luminosity expressed in units of $10^{44}\,\mathrm{erg\,s^{-1}}$. The evolution of the NLXG LF was also assumed to have the same form as that adopted for the QSOs: namely a $(1 + z)$ power-law evolution in the 'break' luminosity, $L_X^*(z)$:

$$L_X^*(z) = L_X^*(0) \, (1 + z)^k \, .$$

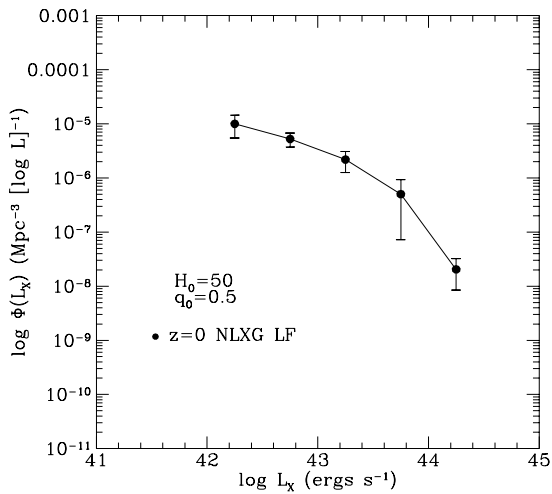

**Figure 4.** The de-evolved X-ray luminosity function for the X-ray-luminous galaxies identified in the EMSS and CRSS.

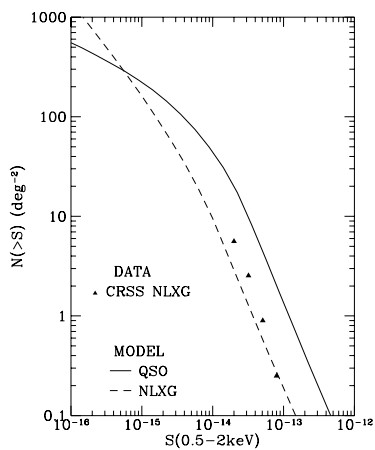

**Figure 5.** The number-flux relation for NLXGs in the CRSS (triangles). The predicted number-flux relation for X-ray galaxies based on the evolution model described in the text is denoted by the dashed line. The solid line illustrates the predicted number-flux relation for QSOs based on evolution model U of Boyle et al. (1994).

Throughout this analysis we adopt a spectral index $\alpha_X = 1$. Although it is slightly harder than the mean spectral index derived for the NLXGs in the CRSS ($\alpha_X = 1.2$, Cilegi et al., in preparation), by using this value we can directly compare our results with those for X-ray-selected QSOs (Maccacaro et al. 1991; Boyle et al. 1993, 1994) also obtained for $\alpha_X = 1$. Nevertheless, the effect of the spectral index of the derived power-law evolution is extremely straightforward, viz. any change to the adopted QSO spectral index, $\alpha_X$, results in an identical change in the derived value of the evolution parameter $k$.

From the maximum likelihood analysis we obtain the following 'best-fit' parameter values: $\gamma_1 = 1.9 \pm 0.25$, $\gamma_2 = 3.8 \pm 0.2$, $L_X^*(0) = 10^{43.21 \pm 0.2} \, \mathrm{erg \, s^{-1}}$, $k = 2.5 \pm 1.0$ and $\Phi^* = 1.6 \times 10^{-7} \, \mathrm{Mpc^{-3}(log \, L^{44} \, erg \, s^{-1})^{-1}}$. This model was an acceptable fit to the data with a KS probability of greater than 10 per cent. Although dominated by a large error (caused by the small redshift range of the sample), the derived rate of evolution for these objects is consistent with that obtained by Boyle et al. (1994) for X-ray QSOs ($k = 3.17$) after the exclusion of the 31 EMSS 'ambiguous' sources (model U in Boyle et al. 1994). A binned version of the de-evolved $z = 0$ LF is plotted in Fig. 4. The observed and predicted number-flux relations for the NLXGs are plotted in Fig. 5, together with the predicted QSO number-flux relation based on model U in Boyle et al. (1994). Both models are integrated out to $z = 4$, with no evolution beyond $z_{max} = 2$. These models predict that the QSOs dominate the X-ray source counts until $S(0.3 - 3.5 \mathrm{keV}) \sim 2 \times 10^{-16} \mathrm{erg \, s^{-1} \, cm^{-2}}$, beyond which NLXGs become the dominant population.

### 3.3 The X-ray background

To estimate the specific intensity due to NLXGs, we integrate over the $0.3 - 3.5$ keV LF:

$$I_{NLXG} = \frac{c}{4\pi H_0} \int_{L_{X_{min}}}^{L_{X_{max}}} \int_{z_{min}}^{z_{max}} \frac{L_X \Phi(L_X)(1 + z)^k}{(1 + 2q_0 z)^{0.5}(1 + z)^{2 + \alpha_X}} \mathrm{d}z \mathrm{d}L$$

where we have chosen our limits to be $L_{X_{min}} = 10^{39} \, \mathrm{erg \, s^{-1}}$, $L_{X_{max}} = 10^{47} \, \mathrm{erg \, s^{-1}}$, $z_{min} = 0$ and $z_{max} = 4$. We use the specific intensity of the extragalactic background in the $1 - 2$ keV energy range derived by Hasinger et al. (1993), $I_{XRB}(1-2 \, \mathrm{keV}) = 1.25 \times 10^{-8} \, \mathrm{erg \, cm^{-2} \, s^{-1} \, sr^{-1}}$. Assuming a spectral index of $\alpha_X = 1$, the specific intensity derived for NLXGs in the $0.3 - 3.5$ keV band is multiplied by 0.28 to obtain the specific intensity in the $1 - 2$ keV band. Note that the fractional contribution of NLXGs to the XRB computed by this method is insensitive to the choice of spectral index, since any increase in the adopted spectral index will be precisely cancelled out in the above equation by the corresponding increase in the evolution parameter $k$ (see above).

For our 'best-fit' model we obtain a specific intensity $I_{NLXG}(1 - 2 \, \mathrm{keV}) = 2.4 \times 10^{-9} \, \mathrm{erg \, cm^{-2} \, s^{-1} \, sr^{-1}}$, corresponding to a contribution to the XRB of 16 per cent. However, this model assumes that the power-law evolution continues unchecked until $z = 4$. A more realistic model would include a maximum redshift $z_{max}$ beyond which the evolution 'switches off' and the comoving space density of QSOs remains constant. Adopting $z_{max} = 2$, consistent with the X-ray evolution of QSOs (Boyle et al. 1993), we derive an overall contribution of 14 per cent. This implies that most of the contribution to the XRB from these galaxies comes from within $z < 2$. The contribution to the XRB from NLXGs with $L_X > 10^{42} \, \mathrm{erg \, s^{-1}}$ (the minimum luminosity observed in our sample) is 6 per cent.

There are also significant uncertainties in the XRB calculation associated with the value of the 'best-fit' parameters. In the model with $z_{max} = 2$ considered here, increasing the evolution parameter to its $1\sigma$ upper limit, $k = 3.5$ results in a contribution of 32 per cent. However, the biggest uncertainty in the estimated contribution to the XRB is the slope of the faint end of the luminosity function. Although the best-estimate is $\gamma_1 = 1.89$, the value of $\gamma_2 = 2$ is within the $1\sigma$ error. With such a steep slope, $I_{\rm NLXG}$ does not converge and the derived contribution to the XRB is dependent on the lower luminosity limit adopted in the integration. For $L_{\rm X_{min}} = 10^{37}\,{\rm erg\,s^{-1}}$, a model with $k = 2.5$, $\gamma_1 = 2.0$ and $z_{\rm max} = 2$, the contribution more than doubles from 14 per cent to 37 per cent. A further decrease of the minimum luminosity in the limit of integration would increase the contribution to even higher values.

## 4 CONCLUSIONS

We have identified 12 X-ray-luminous narrow emission line galaxies (NLXGs) based on a spectroscopic survey of 123 serendipitous $ROSAT$ X-ray sources with $S(0.5 - 2{\rm keV}) > 2 \times 10^{-14}\,{\rm erg\,s^{-1}cm^{-2}}$. The X-ray luminosities of these galaxies range from $10^{42}$ to $10^{43.5}\,{\rm erg\,s^{-1}}$, one to two orders of magnitude greater than observed for typical late-type galaxies. Their optical spectra exhibit emission line ratios characteristic of high ionization, and the sample appears to be composed of approximately equal numbers of starburst and Seyfert 2 galaxies. Coupled with a potentially similar class of objects in the EMSS, the NLXGs exhibit a rate of cosmological evolution, $L_{\rm X} \propto (1+z)^{2.6 \pm 1.0}$, similar to that derived for X-ray QSOs. The NLXG luminosity function exhibits the same two-power-law form as the QSO luminosity function, but with a steeper faint-end slope. Extrapolation of the luminosity function and its evolution yields a predicted NLXG contribution to the $1 - 2\,{\rm keV}$ X-ray background of 15 – 35 per cent, with NLXGs comprising the majority of the faint X-ray sources at $S(0.5 - 2{\rm keV}) < 2 \times 10^{-16}\,{\rm erg\,s^{-1}\,cm^{-2}}$. Further classification of the NLXGs in the EMSS and CRSS samples will enable luminosity functions to be derived for the different types of object that make up this heterogeneous class, but, if the current relative numbers of starburst and Seyfert 2 galaxies observed are typical, then it suggests that both types of object contribute between 7 and 17 per cent of the 1–2 keV background.

## ACKNOWLEDGMENTS

BJB and RGM acknowledge the receipt of Royal Society University Research Fellowships. We are indebted to Dr Mike Irwin for observing the two spectra obtained during spectroscopic service time on the William Herschel Telescope. BJB also acknowledges the support and hospitality of the Smithsonian Astrophysical Observatory. The X-ray data were obtained from the Leicester and Goddard $ROSAT$ archives. This work was partially supported by NASA grants NAGW-2201 (LTSA) and NAS5-30934 (RSDC). The optical spectra were obtained at the William Herschel Telescope at the Observatorio del Roque de los Muchachos operated by the Royal Greenwich Observatory.

## REFERENCES

Avni Y., Bahcall J. N., 1980, ApJ, 235, 694

Baldwin J. A., Philips M. M., Terlevich R. J., 1981, PASP, 93, 5

Boyle B. J., Griffiths R. E., Shanks T., Stewart G. C., Georgantopoulos I. G., 1993, MNRAS, 260, 49

Boyle B. J., Shanks T., Stewart G. C., Georgantopoulos I. G., Griffiths R. E., 1994, MNRAS, in press

Broadhurst T. J., Ellis R. S., Shanks T., 1988, MNRAS, 235, 827

Fabbiano G., 1989, ARA&A, 27, 87

Filippenko A. V., Terlevich R. J., 1992, ApJ, 397, L79

Fruscione A., Griffiths R. E., Mackenty J. W., 1993, in Chincarini D. et al., eds, Observational Cosmology, ASP Conf. Ser. 51. Astron. Soc. Pacif., San Francisco, p.296

Georgantopoulos I. G., Stewart G. C., Shanks T., Griffiths R. E., Boyle B. J., 1993, MNRAS, 262, 619

Georgantopoulos I. G., Stewart G. C., Shanks T., Griffiths R. E., Boyle B. J., 1995, MNRAS, submitted

Giacconi R., Gursky H., Paolini F., Rossi B. B., 1962, Phys. Rev. Lett., 9, 439

Griffiths R. E., Tuohy I. R., Brissenden R. J. V., Ward M. J., 1992, MNRAS, 255, 545

Hasinger G., Burg R., Giacconi R., Hartner G., Schmidt M., Trümper J., Zamorani G., 1993, A&A, 275, 1

Maccacaro T., Della Ceca R., Gioia I. M., Morris S. L., Stocke J. T., Wolter A., 1991, ApJ, 237, 117

McMahon R. G., 1991 in Crampton D., ed., The Space Distribution of Quasars, ASP Conf. Ser. 21. Astron. Soc. Pacif., San Francisco, p.129

Stocke J. T., Morris S. L., Gioia I. M., Maccacaro T., Schild R., Wolter A., Fleming T. A., Henry J. P., 1991, ApJS, 76, 813

This paper has been produced using the Blackwell Scientific Publications TEX macros.